\documentclass[preprint,showpacs,superscriptaddress,floatfix,prl]{revtex4-1}

\usepackage{bm}
\usepackage{amsmath}
\usepackage[dvips]{graphicx}
\usepackage{epsfig}
\usepackage{tabularx}
\newcommand{\ud}{\mathrm{d}}

\begin{document}

\title{Inhomogeneous mechanical losses in micro-oscillators with high reflectivity coating}

\author{E. Serra} 
\affiliation{Interdisciplinary Laboratory for Computational Science (LISC), FBK-University of Trento, I-38123 Povo (Trento), Italy}
\affiliation{Istituto Nazionale di Fisica Nucleare (INFN), Gruppo Collegato di Trento, I-38123 Povo, Trento, Italy}
\author{F. S. Cataliotti}
\affiliation{Dipartimento di Energetica, Universit\`a di Firenze, Via Santa Marta 3, I-50139 Firenze, Italy}
\affiliation{European Laboratory for Non-Linear Spectroscopy (LENS), Via Carrara 1, I-50019 Sesto Fiorentino (FI), Italy}
\affiliation{INFN, Sezione di Firenze}
\author{F.  Marin}
\affiliation{European Laboratory for Non-Linear Spectroscopy (LENS), Via Carrara 1, I-50019 Sesto Fiorentino (FI), Italy}
\affiliation{INFN, Sezione di Firenze}
\affiliation{Dipartimento di Fisica, Universit\`a di Firenze, Via Sansone 1, I-50019 Sesto Fiorentino (FI), Italy}
\author{F.  Marino}
\affiliation{European Laboratory for Non-Linear Spectroscopy (LENS), Via Carrara 1, I-50019 Sesto Fiorentino (FI), Italy}
\affiliation{INFN, Sezione di Firenze}
\affiliation{CNR-ISC, Via Madonna del Piano 10, I-50019 Sesto Fiorentino (FI), Italy}
\author{A.  Pontin}
\author{G. A.  Prodi}
\affiliation{Istituto Nazionale di Fisica Nucleare (INFN), Gruppo Collegato di Trento, I-38123 Povo, Trento, Italy}
\affiliation{Dipartimento di Fisica, Universit\`a di Trento, I-38123 Povo, Trento, Italy}
\author{M. Bonaldi}
\email[Corresponding author: ]{bonaldi@science.unitn.it}
\affiliation{Istituto Nazionale di Fisica Nucleare (INFN), Gruppo Collegato di Trento, I-38123 Povo, Trento, Italy}
\affiliation{Institute of Materials for Electronics and Magnetism, Nanoscience-Trento-FBK Division, I-38123 Povo, Trento, Italy}

\date{\today}

\begin{abstract}
We characterize the mechanical quality factor of micro-oscillators covered by a highly reflective coating. We test an approach to the reduction of mechanical losses, that consists in limiting the size of the coated area to reduce the strain and the consequent energy loss in this highly dissipative component. Moreover, a mechanical isolation stage is incorporated in the device. The results are discussed on the basis of an analysis of homogeneous and non-homogeneous losses in the device and validated by a set of Finite-Element models. The contributions of thermoelastic dissipation and coating losses are separated and the measured quality factors are found in agreement with the calculated values, while the absence of unmodeled losses confirms that the  isolation element integrated in the device efficiently uncouples the dynamics of the mirror from the support system. Also the resonant frequencies evaluated by Finite-Element models are in good agreement with the experimental data, and allow the estimation of the Young modulus of the coating. The models that we have developed and validated are important for the design of oscillating micro-mirrors with high quality factor and, consequently, low thermal noise. Such devices are useful in general for high sensitivity sensors, and in particular for experiments of quantum opto-mechanics.
\end{abstract}

\maketitle


\section{Introduction}

Micro opto-mechanics is earning an increasing interest both for its wide-range applications, including high sensitivity measurements of position, acceleration, force, mass, and for fundamental research. Among the several breakthroughs obtained in the last few years we mention, e.g., radiation-pressure cooling of micro-mochanical oscillators \cite{Cohadon99,Kleckner06,Gigan06,Arcizet06,Schliesser08, grosslacher09},  strong coupling between mechanical and electromagnetic field variables \cite{groblacher09a,verhagen12}, opto-mechanically induced transparency \cite{weis10,safavi11}, and, more recently, the observation of the quantum motion of a nano-mechanical oscillator optically cooled down to its quantum ground state \cite{chan11,safavi12}, closely following similar results previously obtained in the microwave region \cite{oconnell10,teufel11}. Excellent reviews of this field are published, e.g., in Refs. \cite{kippenberg08,marquardt09,favero,aspelmeyer10}.

At present, however, quantum effects have been just observed on the variables of the mechanical oscillator, while important, long-seek quantum properties of light and of the measurement process itself still elude observation. Among them, ponderomotive squeezing \cite{fabre94,mancini94,corbitt,clerk08}, quantum correlations, entanglement and quantum non-demolition measurements \cite{jacobs94,heidmann97,mancini02,vitali}, back-action evasion and measurements reaching and surpassing the standard quantum limit \cite{caves81,jaekel90,arcizet06,clerk08}. On such crucial phenomena, only classical simulations are indeed reported in the literature \cite{marino,verlot10}, besides a recent observation of ponderomotive squeezing induced by a cloud of cold atoms \cite{brooks11}. The main reason must be found in the overwhelming effects of classical noise sources of thermal origin with respect to the weak quantum fluctuations of the radiation pressure.

Most of the conceived experimental schemes cannot take advantage from laser cooling of the main mechanical mode involved in the measurement, while the temperature of the thermal bath is the crucial parameter. For instance, the spectral density of the thermal noise force is proportional to $k_B\,T\frac{m_e\,\omega_0}{Q}$, where $k_B$ is the Boltzmann's constant, $T$ the thermal bath temperature, $m_e$ the oscillator effective mass, $\omega_0/2\pi$ its resonant frequency and $Q$ its quality factor. Therefore, experiments aiming to observe quantum properties of the opto-mechanical interaction benefit from low mass and frequency (on the contrary, for studying the ground state of the oscillator high frequencies are favored by the requirement $\hbar \, \omega_0 > k_B \, T$). Moreover, we remark that the mechanical quality factor plays a crucial role, and that experiments take advantage from high optical quality that allows to build high Finesse cavities. Finally, high  levels of input laser power give stronger radiation pressure, whose fluctuations more easily prevail over thermal noise, provided that the oscillator can be kept at low temperature.

Several kinds of oscillators are currently experimented, such as breathing whispering galleries \cite{Schliesser08,weis10}, thin membranes working as refractive oscillators within high-Finesse cavities \cite{Thompson08,Wilson}, photonic crystal opto-mechanical cavities \cite{chan11,safavi12}. Focusing our interest to Fabry-Perot interferometers with oscillating mirrors, we mention that very low mass oscillators have recently been  conceived and tested, based on free-standing dielectric multi-layer reflectors \cite{Gigan06,cole,cole11}, or on tiny mirrors on shaped thin membranes \cite{grosslacher09,groblacher09a,kuhn11,kleckner11}.

In this article we explore a different approach, focusing on thicker silicon oscillators with high reflectivity coating \cite{marino,arcizet08}. The higher mass is compensated by the possibility to manage high power at low temperatures, thanks to the favorable geometric factor (thicker connectors) and the high thermal conductivity of silicon at cryogenic temperature \cite{glassbrenner}. With such design, it is important to control the mechanical dissipation in the coating layers, in order to exploit the potentially very high $Q$ of silicon oscillators \cite{bib:McGuigan,khine}.    
 Here we report on the losses measured in a micro oscillator covered by a highly reflective coating, and discriminate, by the use of Finite Element analysis (FEM), the contribution of thermoelastic dissipation and coating losses. In particular, we test an approach to the reduction of mechanical losses, where the size of the coated area is reduced as much as possible to reduce the strain in this highly dissipative component and the consequent energy dissipation.
 
The loss angle of an oscillator is defined as:
\begin{equation}
\label{eq:loss}
\phi_{t}=  \frac{\Delta W_{t}}{2 \pi W}
\end{equation}
where $\Delta W_{t}$  denotes the energy dissipated per cycle of vibration
and $W$ denotes the strain energy stored in the elastic body.
Mechanical losses are due to the coupling  between the normal modes and the thermal bath. Several types of losses can be identified, each one associated with a specific dissipation mechanism. The measured loss is the sum of dissipation
arising from different sources $s_i$, so that we can write $\Delta W_{t}= \sum_i \Delta W_{s_i}$. Note that each kind of coupling contributes to the random excitation of the oscillator by drawing from the disordered motion of the bath, called thermal noise. Until recently, experimenters have
focused their attention almost exclusively on homogeneous losses, i.e., losses that are described by an imaginary part of the Young's modulus in a homogeneous body. This is the case for dissipations induced by homogeneously distributed impurities and dislocations. Tools for the evaluation of inhomogeneous distributed losses and thermoelastic losses are now available, based on analytical models or Finite Elements techniques \cite{levin98,yamamoto07,serra}. These developments allow an accurate estimation of the quality factor of real experimental devices.

\section{Setup and characterization}

\subsection{Micro-resonator design and fabrication}
\begin{figure}[t!]
\begin{center}
\includegraphics[width=8.6cm,height=79mm]{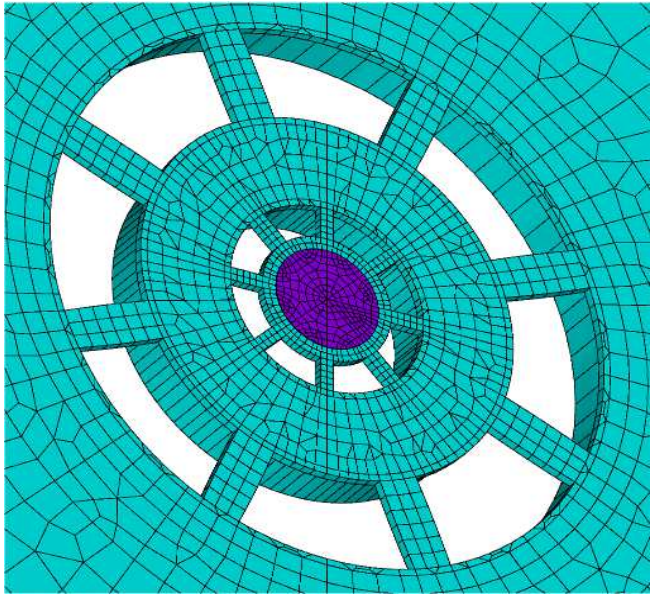}
\end{center}
\caption{Design of the silicon micro-mirror. The diameter of the central oscillating disk is either 700 or 1000 $\mu$m, while the outer diameter of the isolation wheel is 3000 $\mu$m. The beams connecting the central mass with the wheels are either 50 or 100 $\mu$m wide, with length ranging from 300 to 450 $\mu$m, depending by the target frequency of the device. The thickness of the central disk is about 70 $\mu$m, while the thickness of the wheel is 500 $\mu$m.
In the design shown here the optical coating layer (in violet) covers only the central disk; all devices have been produced also in a fully coated version. 
 \label{fig:mesh}}
\end{figure}

Our micro oscillators are fabricated in a silicon-on-insulator  (SOI) wafer where the handle layer is 500 $\rm u m$ while the device layer is 70 $\rm u m$ with a buried oxide of 2 $\mu$m. An example is shown in Fig. \ref{fig:mesh}.
Each oscillator is composed of a central round mass suspended by eight beams. The oscillator is obtained by engraving the handle silicon layer on one side and patterning
the structure in the other side by a Deep-Reacting-Ion Etching (DRIE) process. The buried oxide
is used as etch stop layer.  An outer wheel, fabricated on the full thickness of the wafer, reduces the mechanical coupling between the main oscillator and the high-frequency modes of the whole sample \cite{marino,decumis}. Groups of seven double wheel micro-opto-mechanical oscillators are arranged in $35 \times 35$ mm sectors on the wafer, with slightly different dimensions to cover a frequency range from 160 to 320 kHz, as reported in table \ref{table:data}.  The beams connecting the central mass with the wheels are either 50 or 100 $\mu$m wide, with length ranging from 300 to 450 $\mu$m, and the diameter of the oscillating mirror is either 700 or 1000 $\mu$m.
On the front side of the wafer, a deposition of 17 alternate  $\mbox{Ta}_2 \mbox{O}_5/ \mbox{SiO}_2$  quarter-wave layers for a total thickness of  5.244 $\rm u m$ provides the highly reflective coating (deposited at Advanced Thin Films). We consider two different configurations of the coating: in the first case the coating covers the whole front face of the wafer, including mirror and supporting beams, while in the second case we used a metallic hard mask to coat only the central mirror. The hard mask alignment error was about 100 $\mu$m.

\subsection{Opto-mechanical sensing} 
\begin{figure}
\begin{center}
\includegraphics*[width=8.6cm,height=62mm]{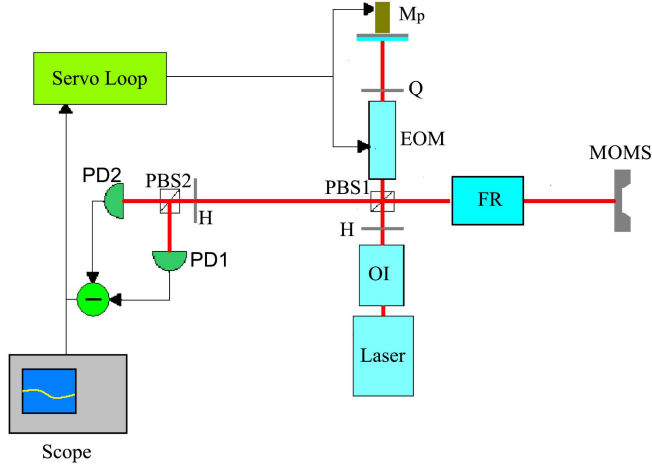}
\end{center}
\caption{Scheme of the experimental apparatus. OI: optical isolator; FR: Faraday rotator;
EOM: electro-optic modulator; H: half-wave plate; Q: quarter-wave plate; PD: photodiode;
PBS: polarizing beam-splitter; MOMS: micro opto-mechanical systems. Lenses and alignment mirrors are not shown in the scheme.}
\vspace{-.3cm}
\label{setup}
\end{figure}

The resonant frequency of each device is determined by the noise displacement spectrum of the mirror's surface.
In order to measure it, we realized a Michelson
interferometer with a balanced homodyne detection scheme (see Fig. \ref{setup}).
The laser radiation is provided by a cw Nd:YAG laser operating at $\lambda$=1064~nm.
After an optical isolator, a polarizing beam-splitter (PBS1) divides the beam into two
parts, orthogonally polarized, forming the Michelson interferometer arms. On the first one (reference beam) an
electro-optic modulator (EOM) and a piezoelectric-driven mirror $M_p$ are used for
phase-locking the interferometer in the condition of maximum displacement sensitivity.
A double pass through a quarter-wave plate rotates by $90^o$ the polarization of the reference beam, which is then reflected by PBS1. The polarization of the beam sent to the micro-mirror (sensing beam), is instead rotated by
a double pass through a Faraday rotator. The sensing beam is focused with a waist of 80 $\mu$m on the coated oscillator, and after reflection it is thus totally
transmitted by PBS1, after which it overlaps with the reference beam reflected by $M_p$.
The overlapped beams are then monitored by homodyne detection. The homodyne detector
consists of a half-wave plate, rotating the polarizations by $45^o$, and a polarizing beam-splitter (PBS2) that divides the overlapped beams
into two equal parts sent to the photodiodes PD1 and PD2, whose outputs are then
subtracted. The signal obtained is a null-average, sinusoidal function of the path difference in the interferometer. Such a scheme is very weakly sensitive to laser power fluctuations.
The difference signal is used as error signal in the interferometer locking servo-loop.
Low-frequency stabilization (up to a few hundred Hz) is carried out by the piezo-mirror
$M_p$, while the wide frequency response of the EOM allows us to extend the locking
bandwidth up to about $30$ kHz. The same signal is sent to a digital
oscilloscope with integrated Fast Fourier Transform analysis, which provides the
noise spectrum. The measured spectrum $S_V$, in V$^2$/Hz, is calibrated through the
expression $S_{xx} = S_V (\lambda/2\pi V_{pk})^2$, where $V_{pk}$ the peak-to-peak value of the interference fringes and $S_{xx}$ is the displacement noise spectrum in m$^2$/Hz.

The maximum record length of our scope (250 kSamples) allows a maximal resolution of 10 Hz (when sampling at 2.5 MSamples/s to avoid aliasing). In order to analyze the narrower resonance peaks, we have also used a digital lock-in amplifier, whose internal local oscillator was tuned at 110 Hz from the interested peaks. The beat note, filtered by the output integrator of the lock-in with a time constant of $640 \mu$s, was then analyzed by the scope with a resolution of 0.1 Hz.

Near the resonance of interest the resonator transfer function $T_x(\omega)$ is well approximated by the harmonic response
\begin{equation}
\label{eq:tx}
T_x(\omega)=  \frac{1}{m_e\left[\omega^2 -\omega_0^2 +i \phi_t \, \omega \omega_0  \right] }
\end{equation}
where  $\phi_t$ is the total loss factor. The associated noise spectrum of thermal origin at the temperature T, is given by the fluctuation dissipation theorem as:
\begin{equation}
\label{eq:sxx}
S_{xx}(\omega)=-\frac{4\,k_B \,T}{\omega}\;\Im m\left[T_x(\omega)\right]= \frac{4\,k_B \,T\phi_t \,   \omega_0}{ m_e \left[ (\omega_0^2-\omega^2)^2+ \phi_t^2 \,\omega^2 \omega_0^2 \right] }
\end{equation}
and its experimental recording can be used to estimate $\omega_0$, $m_{e}$ and $\phi_t$. The value of $m_e$ is then compared with a Finite-Element-Modeling (FEM) estimate to validate the hypothesis of thermal origin for the displacement spectrum.

\subsection{Experimental results}

\begin{figure}[t!]
\begin{center}
\includegraphics[width=8.6cm,height=61mm]{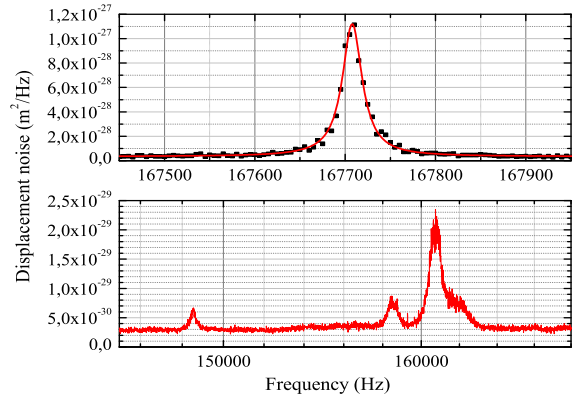}
\end{center}
\caption{ Upper panel: experimental displacement noise spectrum for a micro-mirror homogeneously coated (device D1). The thick solid line (red) is the fitting curve given by Eq. (\ref{eq:sxx}), plus a flat background. Lower panel: displacement noise spectrum obtained from a similar oscillator, but lacking of the outer isolation wheel.
 \label{spettri1}}
\end{figure}

In figure \ref{spettri1} (upper panel) is shown the experimental displacement spectrum for one of the devices (D1 with homogeneous coating). The resonance peak is  
well above the noise floor and allow the estimation of the relevant parameters. The figure also report the fitting curve according to Eq. (\ref{eq:sxx}) (plus a flat background). As a comparison, we also show in the figure (lower panel) the spectrum obtained from a similar oscillator, but lacking of the outer isolation wheel. As already observed for membrane oscillators \cite{decumis}, the coupling with the background silicon modes splits the main oscillator mode into several peaks, increasing the effective mass and reducing the mechanical quality factor.

\begin{figure}[t!]
\begin{center}
\includegraphics[width=8.6cm,height=61mm]{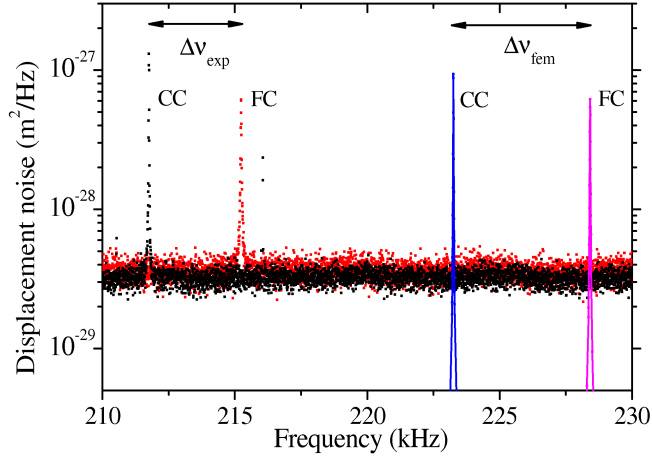}
\end{center}
\caption{Experimental spectra for a micro-mirror resonating at about 215 kHz (device D4). The curves refer to devices with full coating (FC) and central coating (CC). The simulated curves are in agreement within 10\% with the experimental results. The frequency shift $\Delta \nu_{exp}$ and  $\Delta \nu_{fem}$ is due to the different contribution of the coating layer to the structure's stiffness, as discussed in Section \ref{section:Young}.
 \label{fig:spettri}}
\end{figure}

\begin{figure}[t]
\begin{center}
\includegraphics[width=1.0\columnwidth]{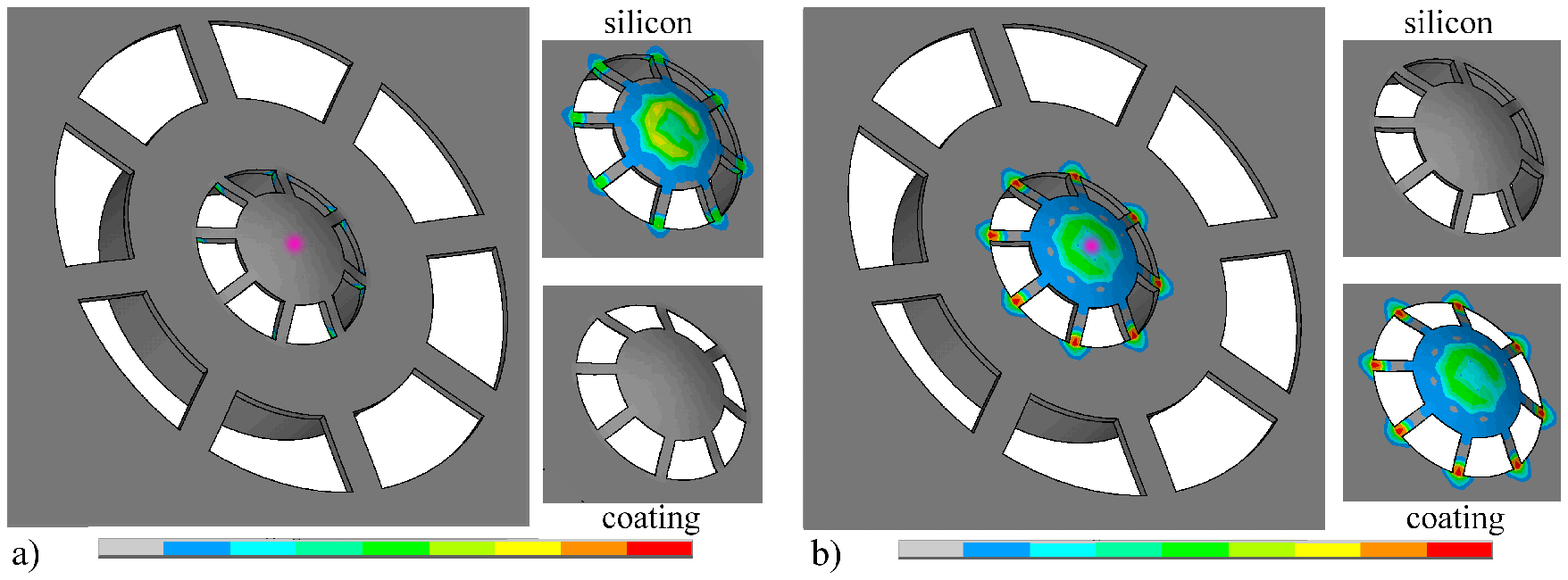}
\end{center}
\caption{Contour plots of the energy dissipated in an oscillation cycle, plotted over the modal shape for the device D8 with full coating: a) thermoelastic loss, b) structural loss. In both cases the insets show separately the dissipation in the silicon wafer and in the optical coating layer. The magenta spot in the center of the mirror gives the size of the optical waist.
 \label{fig:lossesfull}}
\end{figure}

In figure \ref{fig:spettri} we display the experimental and simulated spectra for a micro-mirror resonating at about 215 kHz (device D4). The two experimental curves refer to devices with full coating (i.e., coating over the mirror and the beams; FC) and central coating (i.e., coating just over the mirror; CC).
The resonant frequency in the full coating case is higher, as the coating layer contributes to the stiffness of the supporting beams. This effect  can also be seen in the spectral curves simulated by our Finite-Element model (described below). The effective mass  $m_e$ is evaluated
from experimental data as $(72\pm5)\;\mu$g and $(79\pm5)\;\mu$g, respectively in case of the central coating and full coating. These value are in good agreement with the FEM estimates $(79\pm5)\;\mu$g and $(83\pm5)\;\mu$g, confirming the thermal origin of the noise spectra. As a comparison we note that the physical mass of this mirror can be estimated by its size as 128 $\mu$g; the effective mass is lower because the mirror is not moving as a rigid body, but it bends during the oscillatory motion as shown for instance in figure  \ref{fig:lossesfull}. In table \ref{table:data} we report the parameters measured from experimental data for ten devices. 
We note that the quality factor of each CC oscillators is significantly higher than that of the corresponding FC version. We shall discuss in Section \ref{section:losses} the reason of this behavior and the origin of the observed losses in our devices.

\section{Mechanical losses}
\label{section:losses}
In the model of Eq. \ref{eq:tx} the measured loss factor accounts for all of the dissipative phenomena active in the resonator, as defined in Eq. \ref{eq:loss}. In our case the relevant phenomena are thermoelastic loss and structural loss of the various components of the device, as energy losses through the support are negligible thanks to the isolation wheel. We consider in the following only losses from silicon and optical coating, as the buried silicon dioxide layer is removed during the process from the backside of the mirror and of the beams. 

\subsection{Thermoelastic loss}
 Thermoelastic dissipation was first investigated by
Zener \cite{bib:zener38}: in the presence of a nonzero coefficient of thermal
expansion, when a solid undergoes a vibration other than
pure torsion, the strain field generates a thermal gradient and
thus a heat flow which dissipates elastic energy. This fundamental mechanism sets the loss in MOMS  devices and precision instrumentation at room temperature, for this reason it is the subject of an active area of experimental \cite{Grib,borrielli11}, theoretical \cite{Chandorkar09} and numerical \cite{serra} research.  Only in the case of pure flexure the loss factor can be calculated analytically as $\phi_{Z}=\frac{Y \alpha^2 T}{\rho C_V} \frac{\omega \tau_{Z}}{1+\omega^2 \tau_{Z}^2}$,
where $\alpha$ is the thermal expansion coefficient, $C_V$ the specific
heat per unit volume of the material, $Y$ the Young modulus, $\rho$ the density and $T$ the temperature.  The oscillator thickness  $h$ is involved through the thermal relaxation time 	$\tau_{Z}=\frac{h^2 \rho C_V}{\pi^2 \kappa}$, where $\kappa$ is the thermal conductivity. Even if the real losses depend on the geometry and the anisotropy of the elastic structure and must be evaluated by FEM, these equations give some insight on the behavior of the resonator. For instance, in the case of a silicon cantilever with $h=300\mu$m, we have $\tau_{Z} \simeq 5.6\,\mu$s and the expected loss angle at  250 kHz  is about $2\times 10^{-5}$. This figure limits the Q factor of a silicon flexure to $Q<4\times 10^4$ at room temperature, while better performances could be achieved  at cryogenic temperatures thanks to the changes in the thermal properties of the material \cite{rsizendri}. We also note that at these frequencies the thermoelastic dissipation of the optical coating is negligible, as its average thermal conductivity is 10 times smaller than in silicon \cite{Fejer04} and the thermoelastic heat flow is accordingly smaller.

\subsection{Structural loss}
The structural dissipation is directly related to the imaginary part of the Young's modulus in a homogeneous body, but it cannot be easily evaluated if the device is made of parts with different structural losses. If $\phi_s(\mathbf{r})$ is the loss factor at the position $\mathbf{r}$, the energy $\ud W$ dissipated in one cycle in the volume element  $\ud V$ is simply $E(\mathbf{r})\phi_s(\mathbf{r})\ud V$, where $E(\mathbf{r})$ is the energy stored in the volume element during the motion. In the device the total dissipated energy on one cycle is 
\begin{equation}
\Delta W_{s}= \int E(\mathbf{r})\phi_s(\mathbf{r})\ud V\,\, , 
\label{Ws}
\end{equation}
where both the energy density and the loss factor depend on the position. As a consequence the total loss depends on the shape of the displacement within the resonator: modal shapes involving large strain in more dissipative parts imply higher losses than modal shapes where the same parts are less strained.

\subsection{Finite-Element models}
From the above discussion, it is clear that the dissipative contribution from different phenomena can be identified only by the comparison with simulations produced by a FEM software (in our case, ANSYS Multiphysics). The Finite Element model of the device is based on a three-dimensional 20-node solid element, and  the mechanical response of the mirror+beams structure is evaluated when it is driven by an harmonic pressure over the surface of the mirror. The coupled-field thermoelastic analysis is repeated in the frequency range of interest and the convergence of the results is checked against the mesh density.
In a harmonic thermoelastic analysis, the energy $\ud W$ lost per cycle in an element with volume $\ud V$ is related to the irreversible entropy produced per cycle at the temperature $T_0$  \cite{serra}:
\begin{equation}
 \ud W_{te}= \frac{\ud V}{T_0} \oint_{cycle}   \frac{\partial\theta}{\partial x_i} \ k_{ij} \ \frac{\partial\theta}{\partial x_j} \ud t
\label{eq:entropyLost}
\end{equation}
where $\theta$ is the temperature field, $k_{ij}$  the thermal conductivity tensor
and $x_i$ the spatial coordinates; repeated indexes are summed over.
This quantity was evaluated for each element of the model and integrated over the volume of the resonator to obtain  the expected loss factor $\phi_{te} =\Delta W_{te}/(2 \pi W)$.
We point out that, to be consistent with the measurements, the spatial distribution of the applied pressure is the same as the laser beam intensity profile (Gaussian shape with a waist of 80 $\mu$m), and the resulting displacement of the mirror's surface is weighted by the same Gaussian profile.
The model can simulate, separately or in a cumulative way, both kinds of dissipation under study.
The thermoelastic loss is evaluated from the material properties with no free parameters.
On the contrary the evaluation of structural dissipation requires as input the loss angles of silicon and of the optical coating.

The dissipation of silicon at room temperature is strongly dependent on the size of the sample, mainly due to thermoelastic losses and surface effects. When the thermoelastic loss can be made negligible, the underlying structural loss of silicon wafers is found well below a value of $\phi_{Si}=10^{-6}$ \cite{bib:McGuigan,khine}, that we take as reference in our calculations. We point out that our results will be almost independent from this number, as losses from other parts of the resonator are orders of magnitude higher. Indeed loss angles in the range $\phi=(3-6)\times 10^{-4}$ \cite{yamamoto06,Crooks} represent the state-of-the-art for optical coatings like the one on top of our silicon surface, in spite of the large amount of theoretical and experimental developments carried on by the scientific community interested in gravitational wave detectors.

For each configuration of the coating layer, the loss angle is obtained from FEM data using the following procedure:
\begin{itemize}
\item estimation of the resonant frequency of the mode under study;
\item simulation of the transfer function with a given homogeneous structural loss ($\phi_t=10^{-6}$) and evaluation of resonant frequency $\omega_0$ and effective mass  $m_e$ from the imaginary part of the transfer function, according to equation \ref{eq:sxx};
\item simulation of the transfer function with inhomogeneous losses and evaluation of the resulting loss angle $\phi_s$: FEM data are fitted with the curve reported in equation \ref{eq:sxx}, with given $\omega_0$ and $m_e$ and the loss angle as the free parameter;
\item simulation of the transfer function with thermoelastic loss and evaluation of the resulting loss angle $\phi_{te}$. FEM data are fitted with the curve reported in equation \ref{eq:sxx}, with given $\omega_0$ and $m_e$ and the loss angle as the free parameter.
\end{itemize}
In these simulations we used standard values for the mechanical parameters of silicon and the  value of $Y_c\simeq 74\,$GPa for the Young modulus of the coating, measured as described in Section \ref{section:Young}. We used for the loss angle the value  $\phi_c=6\times 10^{-4}$, within the range of the values measured for this kind of coating in a few experiments at room temperatures \cite{yamamoto06,Crooks}.

\begin{figure}[t!]
\begin{center}
\includegraphics[width=8.6cm,height=61mm]{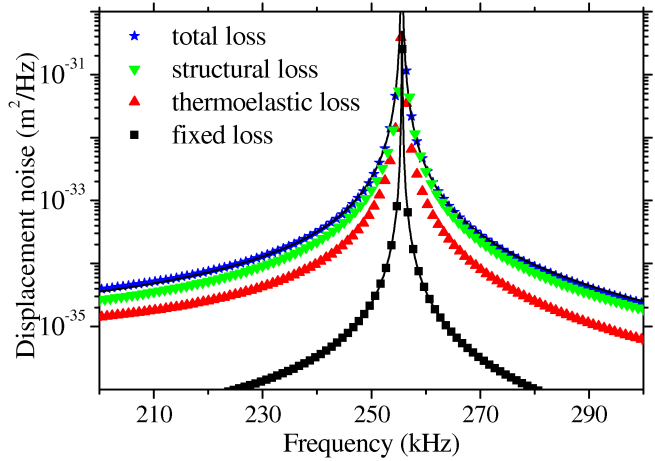}
\end{center}
\caption{Simulated noise spectra for a single device (D8 full coating) with different kinds of losses. Fit with theoretical spectra from Eq. \ref{eq:sxx} is shown only for the calibration curve with fixed loss ($\phi=10^{-6}$) and for the curve with both thermoelastic and structural losses.
 \label{fig:femdata}}
\end{figure}

In figure \ref{fig:femdata} we show the simulated noise spectra for a single device (D8 full coating) with different kinds of losses. The agreement with the Eq. \ref{eq:sxx} is generally good and we estimate in about $\pm$5\% the uncertainty in the evaluation of the loss angle. This error is mainly determined by the accuracy limit  in the measurements of the dimensions of the device and by the discretization errors of the Finite Elements model. The total loss angle is obtained as $\phi_t=\phi_s+\phi_{te}$ and the expected quality factor is $Q=1/\phi_t$. 

To give a better insight on the dissipative behavior of different parts of the device, we plot over the displacement the density of the energy dissipated in an oscillation cycle by thermoelastic and structural losses. In the case of thermoelastic loss we plot the dissipated energy $\Delta W_{te}$ from Eq. \ref{eq:entropyLost} (further integrated over the volume), while in the case of structural loss we plot $\Delta W_s$ given by Eq. \ref{Ws}. In figure \ref{fig:lossesfull} we display the behavior of the oscillator in the case of full coating,
while figure \ref{fig:lossescentral} highlights the energy dissipated in a micro-mirror with  central coating.
These plots show at a glance that thermoelastic loss is mainly due to the flexure of the supporting beams and of the mirror itself, while structural loss is due mainly to dissipation in the coating layer.
For this reason losses are expected to be significantly smaller in the devices with the central coating.

\subsection{Comparison with experimental results}
In table \ref{table:Q_fem} we show the results of the simulations compared with the experimental data for a few devices resonating at different frequencies spanning over our full range (165-325) kHz. The 
agreement between simulation and data is within 20\% in the case of central coating and about 30\% in  the case of full coating. It is remarkable that such agreement is obtained by adjusting a single parameter, namely the loss angle of the coating, within its typical range. On the contrary, in the literature, 
the numerical agreement between simulations and data in dissipative micro-systems is usually obtained by fitting the data with a constant amount of energy loss, that results to be either of just the same magnitude \cite{duwel03} or even about 10 times bigger \cite{cole11} than the simulated losses. This extra-loss accounts for some unmodeled energy leakage through the support system. 

Therefore we can say that our results confirm the validity of our model based on thermoelastic and structural inhomogeneous losses. Our main results can be summarized as follows:\\
a) the outer isolation wheel efficiently uncouples the dynamics of the mirror from  
the support system;\\
b) the reduction of the coated surface allows to reduce the overall loss in the device;\\
c) in the case of central coating the loss is equally contributed by the thermoelastic dissipation in the silicon structure and by the coating's structural loss. 

\section{Evaluation of the Young modulus of the coating}
\label{section:Young}

The availability of the measurements on two sets of identical devices, only differing in the extension of the surface covered by the coating, suggests a way to estimate some structural properties of the coating itself. Indeed, the coating layer do not only contributes to determine the energy loss as described in section \ref{section:losses}, but also, through its mass and its elastic modulus, it affects the dynamical properties of the resonator. 
Specifically, in the oscillators with full coating the deposited oxide layers cover the beams supporting the mirror, thus increasing their stiffness with a negligible increase in the total moving mass, which is mainly determined by the mirror. For this reason the FC devices resonate at a frequency higher than CC devices (table \ref{table:data}), as we indeed observed in the simulations (an example is shown in figure \ref{fig:spettri}).

We point out that the oscillation frequency of each device is a function of all of its structural and geometric parameters, therefore in principle it could be used to obtain the structural parameters of the coating. Unfortunately a typical FEM procedure can evaluate the frequencies of the normal modes with a reproducibility not better that 2\%, due to systematic errors related to the mesh choice and possible biases of the numerical method used by the FEM engine. In our case an error of 2\% in the resonant frequency corresponds to a few kHz, a figure sometimes larger than the observed frequency shift. 

On the contrary the \textit{difference} $\Delta \nu=\nu_{FC} - \nu_{CC}$ is weakly dependent on the choice of the mesh and the numerical method, provided that they remain the same in the evaluation of both  $\nu_{FC}$ and $\nu_{CC}$. Therefore the frequency shift $\Delta \nu$ is mainly dependent on the Young modulus of the coating and on possible thickness differences among the two sets of devices (FC and CC). We can assume that the actual values of density and Young modulus of the coating layer are the same for all of the devices, as they were produced in a single lot. Also the dimensions of the devices in the wafer plane are well determined, as they were obtained by a photo-lithographic process followed by a DRIE etching. On the other hand, the thickness of each wafer depends on the cutting and polishing processes performed on each SOI wafer during the production. 

These considerations can be hardly described analytically, as our devices are not simple geometrical structures, but by FEM simulations we can find for each device phenomenological relations between the frequency shift and the relevant parameters. If $Y_c$ is the Young modulus of the coating, $\Delta h=h_{FC}-h_{CC}$ with $h_{FC}$, $h_{CC}$ the thickness of the FC and CC device layers, we can write 
\begin{equation}
\label{eq:Deltaf}
\Delta \nu = K_Y Y_c + K_h \Delta h + K_m
\end{equation} 
where   $K_Y$, $K_h$, $K_m$ are constants, measured respectively in  Hz/GPa, Hz/$\mu$m and Hz, obtained by the fit of a number of FEM simulations. Here $K_Y$ models the sensitivity of $\Delta \nu$ to changes of the Young modulus of the coating, $K_h$ to changes in the thickness of the device and $K_m$ describes the effect of the mass of the coating layer covering the supporting beams. The actual values of the constants depend on the geometry of the specific device, therefore we have a set of ten equations like Eq. (\ref{eq:Deltaf}) (one for each device considered). If the thickness of the silicon device layers are known, from these equations we can estimate the Young modulus of the coating. 

The $K$ constants that we have found with FEM are reported in table \ref{table:deltaf}. Our Finite-Element models use for the coating a density $\rho_c=4525\;$kg/m$^3$ and a Poisson modulus $\sigma_c=$0.2. The density $\rho_c$ has been calculated from the densities of the Ta$_2$O$_5$ \cite{flaminio} and SiO$_2$ \cite{martin} layers, given that our coating is made of 17 pairs of  Ta$_2$O$_5$/SiO$_2$ layers, which amount to 2184 nm of Ta$_2$O$_5$  and  3060 nm of SiO$_2$.

\begin{figure}[t!]
\begin{center}
\includegraphics[width=1.0\columnwidth]{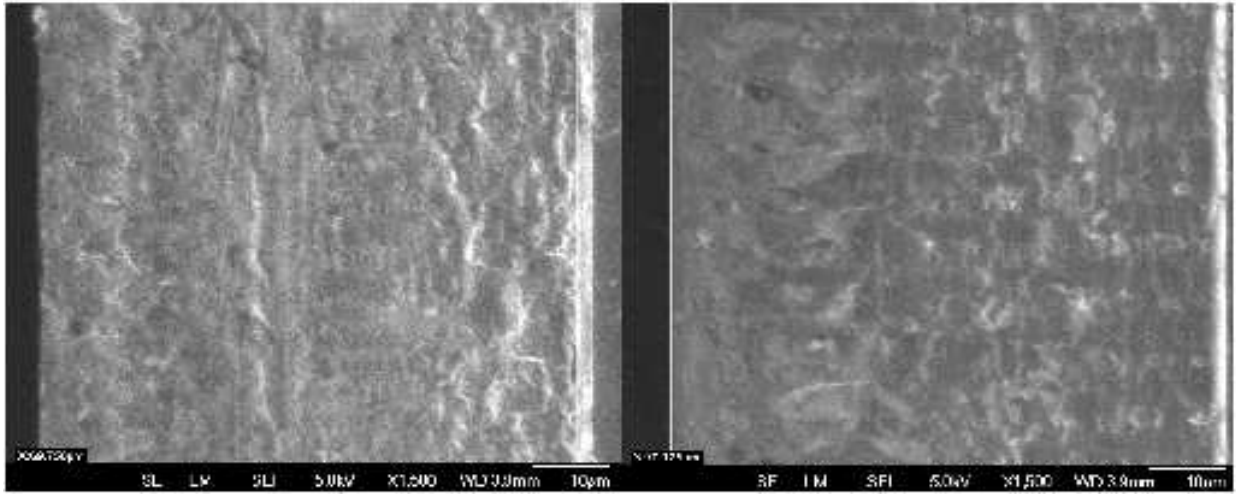}
\end{center}
\caption{Device layer images, including the buried oxide (white region), for the  central coating (left) and full coating (right) SOI samples. These images are used for the evaluation of the thickness variations.  
 \label{fig:SEM}}
\end{figure}

From our data it is clear that, as $K_h$ is on average $\sim2$ kHz/$\mu$m and the frequency shift is on average $\sim4$ kHz, the thickness difference $\Delta h$ must be measured with an uncertainty better than 0.5$\,\mu m$ to allow a 25\% estimate of $\Delta \nu$. Therefore the thickness of the device layers $h_{FC}$ and $h_{CC}$ must be measured with a precision better than 1\%.
The thickness was measured on samples obtained by of the actual wafers used for the production of each series of devices. Two samples (one FC and one CC) having dimensions 1 \rm{cm} x 1 \rm{cm} were cut with a quartz saw from the SOI wafers and then clamped and aligned in the same holder in order to avoid measurement errors due to the reproducibility of the positioning in the SEM's holder. We have used a Jeol JSM-7401F SEM with a magnification of 1500x to have a full view of the device layer and the buried oxide. 
In figure \ref{fig:SEM} the two samples are shown. A white region where electrons are scattered due to the oxide layer is shown on the left side of the two images.  All pictures were taken at a the same tilt angle of the electron beam of $3\,^{\circ}$ (to reach the same experimental conditions). The thickness of the samples were evaluated as an average over measurements performed from the top to the bottom of the images. In this way we could reduce the effect of the surface irregularity near the interface between the device layer and the oxide, which do not give thickness variations greater than 0.2 $\mu \rm{m}$. We have obtained $h_{FC} = 70.1 \pm 0.2 \mu$m and $h_{CC} = 69.75 \pm 0.1 \mu$m. 

Once the evaluated $K$ constants and the measured $\Delta h$ are inserted in equation \ref{eq:Deltaf}, we obtain a set of ten equations for the shift $\Delta \nu$ as a function of $Y$. A comparison with the experimental data, using a best fit approach, allows us to finally obtain for the Young modulus of the coating $Y_c=74 \pm 20\,$ GPa. In table \ref{table:deltaf} we also report for each device the estimated $\Delta \nu_{{fem}}$, obtained with the best fit of $Y_c$, compared with the corresponding measured values $\Delta \nu_{{exp}}$.

\section{Conclusions}
We have measured the mechanical losses in a micro oscillator covered by a highly reflective coating made of 17 pairs of  Ta$_2$O$_5$/SiO$_2$ layers. These devices have been developed to detect radiation pressure coupling between a low-mass moving mirror and an incident light field in a Fabry-Perot cavity, an application where it is crucial to reduce as much as possible the thermal noise. Our approach exploits silicon micro-oscillators built in the device layer of a SOI wafer. Their design include an embedded isolation stage to limit the mechanical coupling to the background wafer. With respect to other kinds of oscillating mirrors, recently proposed and tested, that use much thinner structures for the mirror suspension and even the oscillating mass, our devices suffer from a higher effective mass. This drawback is compensated by the possibility to use high laser power at low temperature, thanks to both the favorable geometric factor (thicker links) and the high thermal conductivity of silicon at cryogenic temperature \cite{glassbrenner}. For instance, at 4.5 K our oscillators can bear several mW of absorbed power keeping the temperature spread within the device below 1 K. Considering that typical absorption in high reflectivity coatings is of few ppm, the intracavity laser power can reach $\sim 1$ kW (obtainable with the typical Finesse of $\sim 30000$ and an input power of $\sim 100$ mW). We remark that temperature homogeneity is important to avoid effects of non-equilibrium thermal noise \cite{NEnoise}.   

The devices that we have presented are also characterized by a good sturdiness, useful for possible applications to high sensitivity sensors. For instance, we can clear them with standard cotton-tipped sticks to routinely obtain optical cavities with Finesse around 40000, using a standard 100 ppm transmission input mirror.

In order to optimize the efficiency of the micro-mechanical oscillators in quantum optics experiments, it is important to design devices with reduced mechanical dissipation. In this work we analyze in particular the advantage brought by a reduced coating area. The results
 are discussed on the basis of an accurate analysis of the possible losses and validated by a number of FEM simulations. Thanks to the use of two families of devices (full coating and central coating), the contribution of thermoelastic dissipation and coating losses could be separated and the observed change in quality factor is in agreement with the expected value, demonstrating that the reduction of the coated surface allows to reduce the overall loss in the device. We also note that the absence of unmodeled losses confirms that the outer isolation wheel efficiently uncouples the dynamics of the mirror from the support system.

\begin{figure}[t]
\begin{center}
\includegraphics[width=1.0\columnwidth]{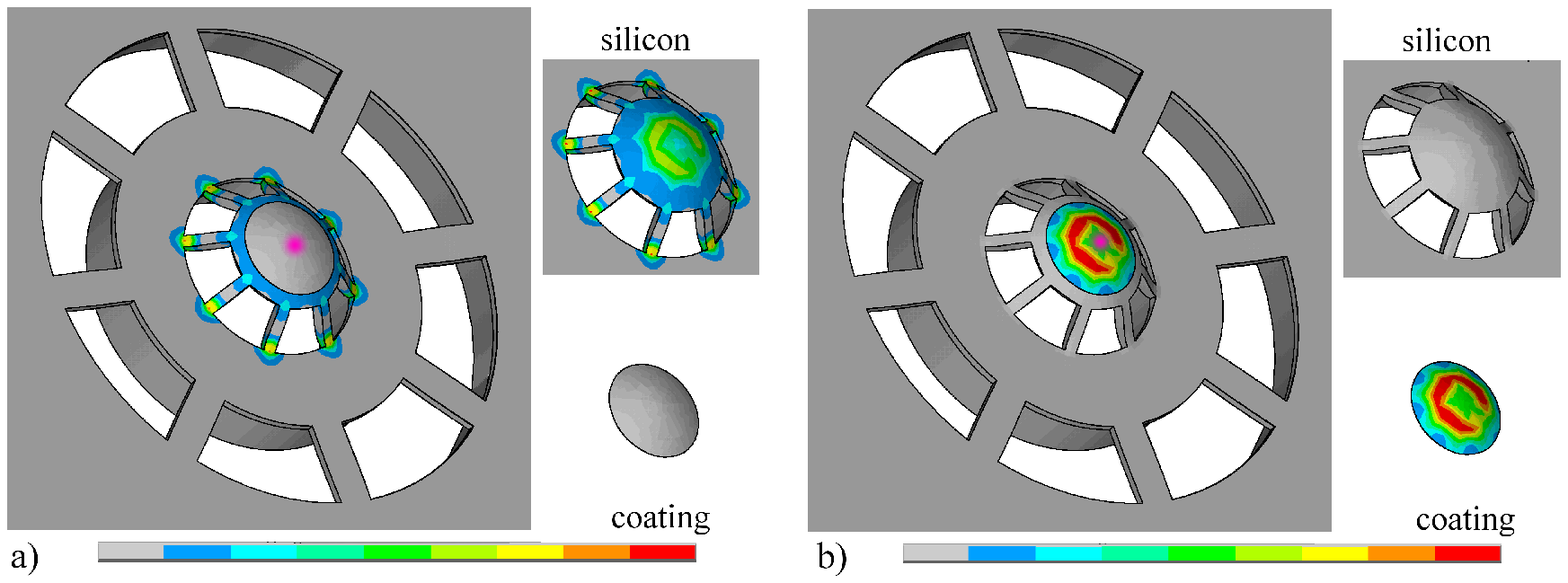}
\end{center}
\caption{Contour plots of the energy dissipated in an oscillation cycle, plotted over the modal shape for the device D8 with central coating: a) thermoelastic loss, b) structural loss. In both cases the insets show separately the dissipation in the silicon wafer and in the optical coating layer. The magenta spot in the center of the mirror gives the size of the optical waist.
 \label{fig:lossescentral}}
\end{figure}

The resonant frequencies evaluated by FEM are also in good agreement with the experimental data, within the reproducibility observed by measuring different samples of the same device. By comparing the resonant frequency of full coating and central coating devices, we could estimate of the Young modulus of the coating, in agreement with measures performed on completely different systems. Indeed, the coating applied over our mirrors is the same currently used in optical cavities for metrology experiments and in the large mirrors of gravitational wave detectors. 
As a final remark we observe that thermoelastic loss could be reduced well below 10$^{-6}$ by cooling the system at liquid helium temperatures \cite{rsizendri}, where therefore our devices should reach a quality factor exceeding 3$\times 10^4$, at the state of the art for micro-mirrors in our frequency range. On the other hand our design could be further improved, as most of the losses at low temperature occur in the coating covering the mirror, due to the bending of the mirror itself (see figure \ref{fig:lossescentral}). A design where only the beams bend during the oscillation, while the mirror moves as a solid body with negligible strain,  could be very useful to further reduce the energy loss in the coating. The FEM models described in this work and validated by the comparison with the experiment represent an essential tool for such designing activity. The requirements proposed, e.g., in Ref. \cite{marino} for obtaining pondero-motive squeezing (in particular, a Q of $\sim 10^5$ in an oscillator with a mass of 50$\;\mu$g and a frequency of 100 kHz operating at liquid helium temperature) are achievable by such design.

\newpage

\begin{table}
\caption{Parameters of the different micro-mirrors as obtained by the fit of experimental thermal noise spectra with the expression (\ref{eq:sxx}).  The experimental errors are estimated from the reproducibility of the data obtained with different samples of the same device.\label{table:data}}
\footnotesize
\begin{tabular}{l ccc ccc ccc}
\hline
\hline
  &\multicolumn{3}{c}{Oscillator size}&\multicolumn{3}{c}{Full coating}&\multicolumn{3}{c}{Central coating}\\
\hline
\hline
 & Central disk  &Beam  &Beam  &Freq. (kHz)  & $m_e$($\mu$g)   &Q&Freq. (kHz) { }&{ } $m_e$ ($\mu$g)&Q\\ 
 & diameter ($\mu$m) & width ($\mu$m)& length ($\mu$m) &$\pm$ 1\% & $\pm$ 10\% & $\pm$ 10\% & $\pm$ 1\% &  $\pm$ 10\%& $\pm$ 10\%\\
 \hline
D1 & 1000&100&400&167.{\scriptsize 7} & 82&6300  &164.{\scriptsize 7}&76&16300 \\
D2 & 700&100&300&324.{\scriptsize 7} &49 &5200  &317.{\scriptsize 3}&50&13400 \\
D3 & 700&100&350&279.{\scriptsize 0} & 59& 6000  &273.{\scriptsize 1} &47& 16200\\
D4 & 1000&100&300&215.{\scriptsize 2} &79& 6500 &211.{\scriptsize 7} & 72& 17300\\
D5 &700&100&400&242.{\scriptsize 3} &59& 6200 &238.{\scriptsize 0} &46& 19700\\
D6 &700&50&400&208.{\scriptsize 1} &45& 5200 &203.{\scriptsize 3}&47&17200\\
D7 &700&50&450&181.{\scriptsize 5} &50& 6100  &177.{\scriptsize 4} &47& 16700\\
D8 &1000&100&250&249.{\scriptsize 4} & & 5900  &242.{\scriptsize 1} & & 15100\\
D9 &1000&100&200&288.{\scriptsize 9} & & 4100  &280.{\scriptsize 5} & & 11900\\
D10 &1000&100&150&338.{\scriptsize 2} & & 4900  &328.{\scriptsize 6} & & 8200\\
\hline
\hline
\end{tabular}
\end{table}

\newpage

\begin{table}
\caption{Comparison between experimental and simulated Q values.
\label{table:Q_fem}}
\footnotesize
\hspace*{3mm}
\begin{tabular}{@{}lcccccccc}
\hline
\hline
  &\multicolumn{4}{c}{Full coating}&\multicolumn{4}{c}{Central coating}\\
\hline
 & $\phi_s$ &  $\phi_{te}$  &Q$_{FC}$  &Q$_{exp}$&$\phi_s$&  $\phi_{te}$  &Q$_{CC}$  &Q$_{exp}$\\
  &$\pm$  5\% & $\pm$ 5\% &  $\pm$ 10\%  &  $\pm$ 10\% & $\pm$ 5\% &  $\pm$ 5\%& $\pm$ 10\% &   $\pm$ 10\%\\
\hline
D1 { }&6.2$\times 10^{-5}$ & 4.5$\times 10^{-5}$ & 9300&6300  & 2.3$\times 10^{-5}$ & 4.5$\times 10^{-5}$  & 14800&16300 \\
D2 { }&7.4$\times 10^{-5}$ & 2.8$\times 10^{-5}$  & 9700 &5200  &3.4$\times 10^{-5}$ & 2.8$\times 10^{-5}$  & 16200&13400 \\
D4 { }& 7.1$\times 10^{-5}$ & 3.7$\times 10^{-5}$ & 9200& 6500 &3.2$\times 10^{-5}$ &3.7$\times 10^{-5}$  &  14400 & 17300\\
D5 { }& 7.1$\times 10^{-5}$&  3.4$\times 10^{-5}$ & 9500& 6200 & 3.0$\times 10^{-5}$ &  3.4$\times 10^{-5}$  & 15600& 19700\\
\hline
\hline
\end{tabular}
\end{table}

\newpage

\begin{table}
\caption{Proportionality constants $K$ describing the dependence of $\Delta \nu$ from the relevant parameters, according to equation \ref{eq:Deltaf}. $K_E$ is the sensitivity  to changes of the Young modulus of the coating, $K_h$ to changes in the thickness of the device and $K_m$ describes the effect of the mass of the coating layer covering the supporting beams. The constants are evaluated by fits on the results of FEM simulations.  In the last two columns we report the expected values of the frequency shift $\Delta \nu_{{fem}}$, and the corresponding experimental measurements $\Delta \nu_{{exp}}$. In the evaluation of $\Delta \nu_{{fem}}$ we have used $\Delta h=0.4 \pm 0.3\; \mu $m, directly measured, and the best fit evaluation $Y_c=74 \pm 20\,$ GPa.
\label{table:deltaf}}
\begin{tabular}{lrllll}  
\hline
\hline
   & $K_E$ \ \ \ \ \ \ &\ \ $K_h$  \  \ & \ \ \ \ $K_m$  &$\Delta \nu_{{fem}}$ & $\Delta \nu_{{exp}}$\\
\hline
 & Hz/GPa \ \ & Hz/$\mu$m  \ \ & \ \ \ \ Hz &  kHz& kHz \\
\hline
\hline
D1 &78.6 $\pm$ 0.1 \ \ & 2258$\pm$3  & -2845 $\pm$3  \ \ &3.{\scriptsize 8} $\pm$ 2 \ \ & 3.{\scriptsize 0} $\pm$ 3\\
D2 &105.8 $\pm$ 0.3 \ \ & 4405$\pm$6  & -1826 $\pm$4  &7.{\scriptsize 6} $\pm$ 2   & 7.{\scriptsize 5} $\pm$ 4\\
D3 & 110.2 $\pm$ 0.2 \ \ & 3868$\pm$6  & -2130 $\pm$4 &7.{\scriptsize 5} $\pm$ 3   & 6.{\scriptsize 7}  $\pm$ 2\\
D4 & 83.5 $\pm$ 0.1 \ \ & 3007$\pm$4  & -2885 $\pm$9 &4.{\scriptsize 4} $\pm$ 3   & 3.{\scriptsize 5} $\pm$ 3\\
D5 & 92.2 $\pm$ 0.2 \ \ & 3399$\pm$5  & -2108 $\pm$3  &6.{\scriptsize 0} $\pm$ 3   &6.{\scriptsize 2} $\pm$ 2\\
D6 & 97.2 $\pm$ 0.2 \ \ & 2921$\pm$3  & -1004 $\pm$2 &7.{\scriptsize 3} $\pm$ 3   & 6.{\scriptsize 0} $\pm$ 2\\
D7 & 80.6 $\pm$ 0.1 \ \ & 2581$\pm$3  & -1003 $\pm$2 & 6.{\scriptsize 0} $\pm$ 2 & 4.{\scriptsize 1}  $\pm$ 3\\
D8 & 95.7 $\pm$ 0.2 \ \ & 3129$\pm$5  & -3155 $\pm$4 &5.{\scriptsize 1} $\pm$ 3  &  7.{\scriptsize 3} $\pm$ 4\\
D9 & 102.0 $\pm$ 0.2 \ \ & 3546$\pm$6  & -3171 $\pm$4 &5.{\scriptsize 7} $\pm$ 3  &  8.{\scriptsize 5} $\pm$ 4\\
D10 & 108.7 $\pm$ 0.4 \ \ & 4041$\pm$8  & -3102 $\pm$5 &6.{\scriptsize 5} $\pm$ 3  &  9.{\scriptsize 5} $\pm$ 4\\
\hline
\hline
\end{tabular}
\end{table}

\newpage

\newpage

\newpage

\newpage

\newpage

\newpage

\newpage

\newpage

\end{document}